\documentclass{aa}  
\usepackage{graphicx,natbib}  
\newcommand{\ltsima} {$\; \buildrel < \over \sim \;$}  
\newcommand{\gtsima} {$\; \buildrel > \over \sim \;$}  
\newcommand{\lta} {\lower.5ex\hbox{\ltsima}}  
\newcommand{\gta} {\lower.5ex\hbox{\gtsima}}  
\newcommand{\Ha} {H$\alpha$}  
\newcommand{\Hb} {H$\beta$}  
\newcommand{\cf} {$\Omega_{\rm{CELR}}$}

\newcommand{\ergs}{\>{\rm erg}\,{\rm s}^{-1}}

\defcitealias{gijs02}{VK02}
\defcitealias{chiaberge:ccc}{CCC99}
\begin{document}  

\title{The HST view of the nuclear emission line region in low luminosity
radio-galaxies \thanks
{Based  on observations obtained at
the  Space  Telescope Science  Institute,  which  is  operated by  the
Association of  Universities for Research  in Astronomy, Incorporated,
under NASA contract NAS 5-26555.}} 
  
\titlerunning{The emission line region in low luminosity radio-galaxies}  
  
\author{Alessandro Capetti
\inst{1}
\and    
Gijs Verdoes Kleijn \inst{2}    
\and   
Marco Chiaberge \inst{3}}   
   
\offprints{A. Capetti}  
     
\institute{INAF - Osservatorio Astronomico di Torino, Strada
  Osservatorio 20, I-10025 Pino Torinese, Italy\\ 
\email{capetti@to.astro.it}
\and 
European Southern Observatory, Karl-Schwarzschild-strasse 2, 
D-85748 Garching, Germany,\\
\email{gverdoes@eso.org}
\and
INAF - Istituto di Radioastronomia, Via Gobetti 101, I-40129 Bologna, Italy\\
\email{chiab@ira.cnr.it}}

\date{}  
   
\abstract{We study the properties of  the emission line regions in two
samples  of  low luminosity  radio-galaxies  (LLRG),  focusing on  the
compact emission  line region (CELR)  revealed to be  a characteristic
feature of these objects by HST narrow-band imaging. 
We  find  a  strong correlation  between  line  and  optical
continuum nuclear  emission, suggesting  that the optical  cores (most
likely of non thermal origin) can be directly associated to the source
of ionizing photons, i.e. that we are seeing a jet-ionized narrow line
region.  A  photon budget argument indicates that  the optical nuclear
sources produce  a sufficient photon  flux provided that  the covering
factor of the circum-nuclear gas is rather large, on average $\sim 0.3$.
Analysis of  HST images and  spectra suggests that
the CELR may take the form of a pc-scale, high filling factor,
structure, possibly an optically thin torus.
Estimates  of the CELR  mass lead  to values  as small  as 10  - 10$^3
M_{\sun}$ and photon counting sets a limit to the BLR mass of $M_{BLR}
< 10^{-2} M_{\sun}$.  When  considered together with the low accretion
rate and the tenuous torus structure,  a general paucity of gas in the
innermost  regions   of  LLRG  emerges  as   the  main  characterizing
difference from more powerful AGN.

}
\maketitle  
  
\section{Introduction}  

The  study of  low luminosity  active galactic  nuclei  (LLAGN) (i.e.,
bolometric luminosity typically \ltsima $10^{42} \ergs$) is a key tool
to  improve  our  understanding  of  the  process  of  accretion  onto
super-massive black  holes. The  LLAGN provide us  with a view  of the
physical mechanisms  related to  mass accretion in  a regime  which is
complementary to (and  possibly different from) those at  work in high
luminosity AGN (HLAGN).   The LLAGN represent a link  between the high
luminosity AGN and the population of quiescent galaxies, as it is now
widely   recognized   that   most   galaxies  host   a   super-massive
black-hole. Exploring the different manifestations of nuclear activity
across the largest possible range of luminosity is then a crucial step
towards a  better understanding of the connections  between active and
non-active galaxies.

LLAGN  are  found  in bright  galaxies  at  a  much higher  rate  than
HLAGN. In a  sample of nearby galaxies studied  by \citet{ho97} $\sim$
30  \% show  a  measurable level  of  nuclear activity,  based on  the
presence of emission lines  of non-stellar origin. Recently, this high
fraction have  been confirmed  by analysis of  SDSS spectra of  a much
larger sample  \citep{miller:03}. Radio-loud AGN, which  are the focus
of this paper, are associated almost invariably with bright elliptical
galaxies; their radio-luminosity  function indicates that again $\sim$
30  \%  of   the  galaxies  with  $M_V  <   -21$,  are  associated  to
radio-sources  brighter   than  $10^{29.5}$  erg   s$^{-1}$  Hz$^{-1}$
\citep{auriemma77}.

LLAGN,  both radio-quiet and  radio-loud, then  represent the  bulk of
active galaxies  (and a substantial  fraction of the  overall galaxies
population).  Unfortunately,   their  study  has   been  significantly
hampered by the contamination from host galaxy emission which, in most
observing bands,  dominates the emission  from the AGN.   To constrain
the  physical  processes  at  work  in these  objects  it  is  clearly
necessary to disentangle the AGN and host's contributions via spectral
decomposition  or high  resolution imaging.   In the  last  few years,
thanks to  the HST  and, more recently,  to Chandra imaging,  this has
become routinely possible down to the lowest end of the AGN luminosity
function, at least in the nearby Universe.

An example  of the importance of  this approach comes  from the recent
results  obtained  for  low  luminosity  radio-galaxies  (LLRG).   HST
broad-band imaging of the sample formed by the 3CR radio-galaxies with
FR~I morphology revealed, in the majority of the targets, the presence
of unresolved optical  nuclei \citep[ hereafter CCC99]{chiaberge:ccc}.
Fluxes and luminosities of these  source show a tight correlation with
the radio  cores, extending over  four orders of magnitude.   This has
been interpreted as being due to a common non-thermal emission process
in the  radio and optical band,  i.e.  that we are  seeing the optical
emission from the base of  a relativistic jet.  This interpretation is
supported by evidence for anisotropy  of the optical sources, by their
spectral  indices  and,  when  available,  by  their  spectral  energy
distributions \citep{chiaberge:cena,chiaberge:6251}.  The dominance of
the non-thermal emission and  the corresponding limits to the emission
from  the accretion  disk, might  imply a  profoundly different nuclear
structure  with respect  to classical,  more luminous  AGN.   The high
fraction  (85 \%)  of objects  with detected  optical  nuclear sources
suggests  a  general  lack  of  obscuring molecular  tori,  a  further
distinction with respect to other classes of AGN.

\citet[ hereafter  VK02]{gijs02} reached similar  conclusions from the
observations of another sample of LLRG selected from the UGC catalogue
(see Sect. \ref{sample} for the sample definition).  They also noticed
the characteristic presence of a compact nuclear component of emission
line, in all  but  one case  (namely,  UGC 6635  characterized by  a
complex dusty morphology). Extended emission is also
often detected,  almost always  associated with dust.  Nonetheless the
CELR represents in most cases the dominant component, being a fraction
of at least 10\% of the line flux within 1 kpc, with a median fraction
of 70  \%. They found a  close correlation between the  radio core and
nuclear line emission luminosities.

These observations  set the  stage for a  better understanding  of the
physics of  low luminosity radio-galaxies, focusing  on the properties
of the emission line regions. Here we build on the results obtained by
\citeauthor{gijs02}  by combining the  UGC FR~I  sample with  the FR~I
galaxies  from the  3CR sample  for which  we analyze  HST narrow-band
images and  spectra, thus extending the  coverage in radio-luminosity.
Similarly to the  UGC sample, the 3CR FR~I  sources shows a ubiquitous
presence  of CELRs.   But  the  higher  detection rate  of
optical nuclei in the 3C with respect to the UGC sample, enables us to
perform  a more  detailed  analysis of  the  relationship between  the
different  emission  components in  LLRG.  Aim  of  this paper  is  to
establish the source of ionization  of CELRs in LLRG and to constrain the
physical properties of  the CELRs.

We decided  here to focus our  study on the central  component of line
emission, although additional extended line emission is often present,
as it  represents a  well defined structure  recognizable in  all LLRG
examined,  thus a  characteristic  feature associated  to the  nuclear
activity of these objects.  By  limiting ourselves to the nuclear line
component,  we aim  at minimizing  the contamination  by line-emitting
mechanisms not  directly related  to the active  nucleus (such  as the
extended emission line  disk of M~87, which is  most likely powered by
internal  shocks  \citep{dopita:m87} or  by  clumps  of line  emission
associated to star formation \citep[e.g.][]{odea:3c236}).
 
The paper is organized as  follows: in Sect.  \ref{sample} we describe
the  samples   selected  for  our  analysis  and   the  available  HST
observations.   In  Sect.  \ref{results}  we  study the  relationships
between  the line  luminosities  with other  relevant
properties  of  our  objects  which  are  then  used  to  explore  the
ionization mechanism of the CELR (Sect. \ref{mechanism}), its physical
structure  (Sect.   \ref{physics})  and  the  possible  presence  (and
origin) of  the classical correlation between line  and extended radio
luminosity  in  LLRG (Sect.   \ref{line-vs-radio}).   Our results  are
discussed    in    Sect.    \ref{discussion}   and    summarized    in
Sect. \ref{conclusions}.   A Hubble  constant of H$_{\rm  o}$ =  75 Km
s$^{-1}$ Mpc$^{-1}$ is adopted.

\section{The samples and the HST observations}  
\label{sample}  

For  our   analysis  we  selected   two  samples  of   low  luminosity
radio-galaxies.   The first is  formed by  the 33  radio-galaxies with
FR~I morphology part of  the 3C catalogue\footnote{Here we include NGC
6251  in the  3C sample,  following its  revision  by \citet{laing83};
conversely  3C~84 with its  ill-defined and  peculiar radio-morphology
has been removed  from the original list.}.  The  second is the sample
of 21  radio-bright UGC galaxies defined  by \citet{noelstorr03} (E/SO
galaxies brighter than $m_B < 14.6$ in the optical and than 150 mJy at
1400 MHz, but see their  paper for the precise sample definition). The
3C samples is selected at  low radio frequency and thus is essentially
free  from  orientation  biases.    Concerning  the  UGC  sample,  the
selection at 1.4 GHz does  not warrant an unbiased selection. However,
the radio  cores from VLA  1490 MHz measurements (FWHM  $\sim 1\farcs5
-3\farcs75$) constitute always less than  $22\%$ of the total 1400 MHz
flux and typically $\sim 6\%$ \citep{xu00} indicating that orientation
can only have a marginal effect on the sample selection.

As there are  7 objects in common, the total number  of objects in the
combined sample is 47.  The 3C sample alone spans the range $10^{30.5}
<  {\rm  log}  L_{ext}  [{\rm   W  Hz}^{-1}]  <  10^{33.5}$  in  radio
luminosity, but its lowest end  is only scarcely populated.  Only with
the inclusion  of the  UGC sample  (all UGC sources  have $  {\rm log}
L_{ext} <  10^{32.2}$) the coverage at all  levels of radio-luminosity
is sufficiently uniform to  adequately explore the different behaviour
over the whole luminosity range spanned by LLRG.

For  the whole UGC  sample, in  addition to  broad band  imaging, both
narrow  band   and  medium  resolution  HST   spectra  are  available.
Emission-line measurements of the  Compact Emission Line Region (CELR)
have  been  obtained   from  narrow-band  imaging  by  \citet{gijs99},
isolating the  contribution of the central line component,
as  well   as  from   the  spectra  \citep{noelstorr03}.    Since  the
line-widths in most objects are such that de-blending of \Ha\ from the
[N II] lines cannot be performed accurately from the medium resolution
spectra and  since the  narrow band filters  do not allow  to separate
these lines, we  always measured the fluxes from  the whole \Ha+[N II]
complex.  Measurements from images and  spectra show in general a very
close agreement,  within less  than a  factor of 2,  except for  a few
objects  for which  the  spectroscopic target  acquisition failed;  we
decided to consider, when available, those obtained from spectra.

For nine 3C/FRI sources not  part of VK02 sample, we retrieved all
STIS spectra  and narrow band  images available from  the Multimission
Archive  at   the  Space   Telescope  Science  Institute   (see  Table
\ref{tabobs} for a summary).

All  spectra were  taken as  part of  program GO~8700  with  the G750L
grating   (centered  at   7751\AA,  it   covers  the   spectral  range
5240-11490\AA\  and it  has an  average  dispersion of  4.92 \AA\  per
pixel) in combination with a $0.2^{\prime\prime}$ width slit, centered
on  the unresolved  nucleus  of each  galaxy.   The observations  were
performed in snapshot mode, with a total exposure time of 360 sec (720
s  in  the  case of  3C~78),  splitted  into  two exposures  to  allow
cosmic-ray rejection.  The spectra  were calibrated by the standard On
The Fly Reprocessing (OTFR) system,  but we have further processed the
2-D spectra with the IRAF  task {\it cosmicrays} to eliminate residual
cosmic rays and hot pixels that were not removed by the OTFR.

Narrow-band  images are  available for  6 objects.  Four  targets were
observed as  part of the  3C emission-line snapshot program  (GO 5957)
using the F673N  narrow filter (FR680N ramp filter  for 3C~277.3) with
an exposure time of 600 splitted into two exposures. The remaining two
objects, 3C~442 and NGC~6251, were also observed with the F673N filter
but for different  programs, with an exposure time of  2500 and 5700 s
respectively.   All images  were  calibrated by  the OTFR.   Continuum
removal was obtained  subtracting a scaled version of  the F702W broad
band  image  (F555W for  NGC  6251), appropriately  scaled taking  into
account exposure time and filter sensitivity.

\begin{figure*} \resizebox{\hsize}{!}
{\includegraphics[width=\textwidth]{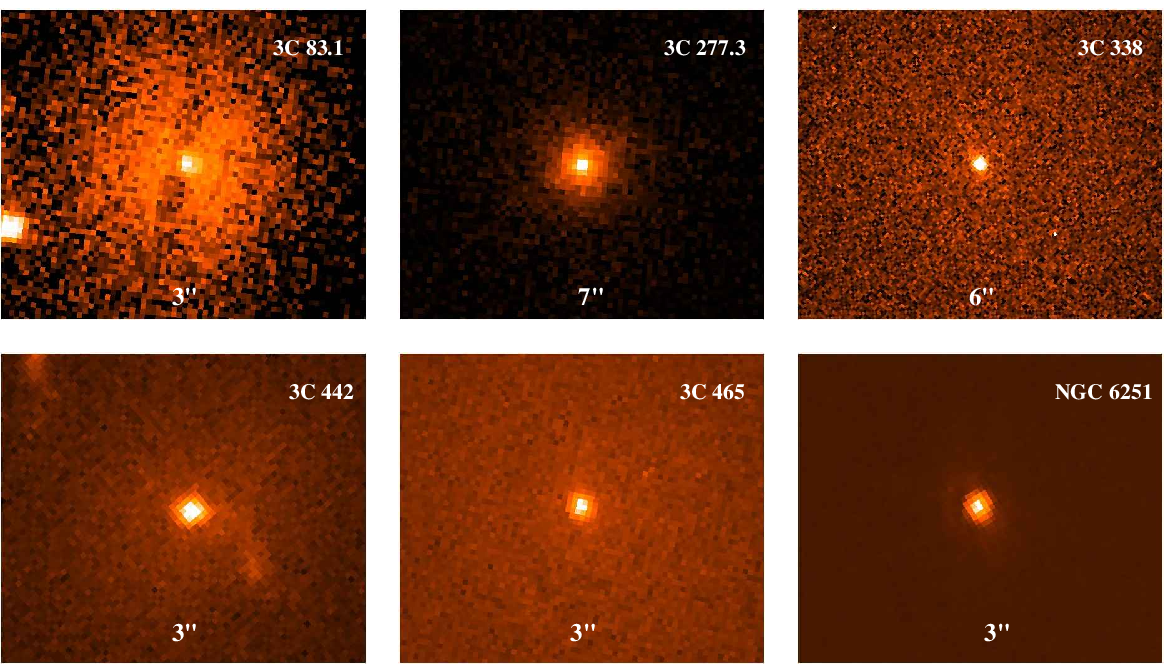}}
\resizebox{\hsize}{!}
{\includegraphics[width=\textwidth]{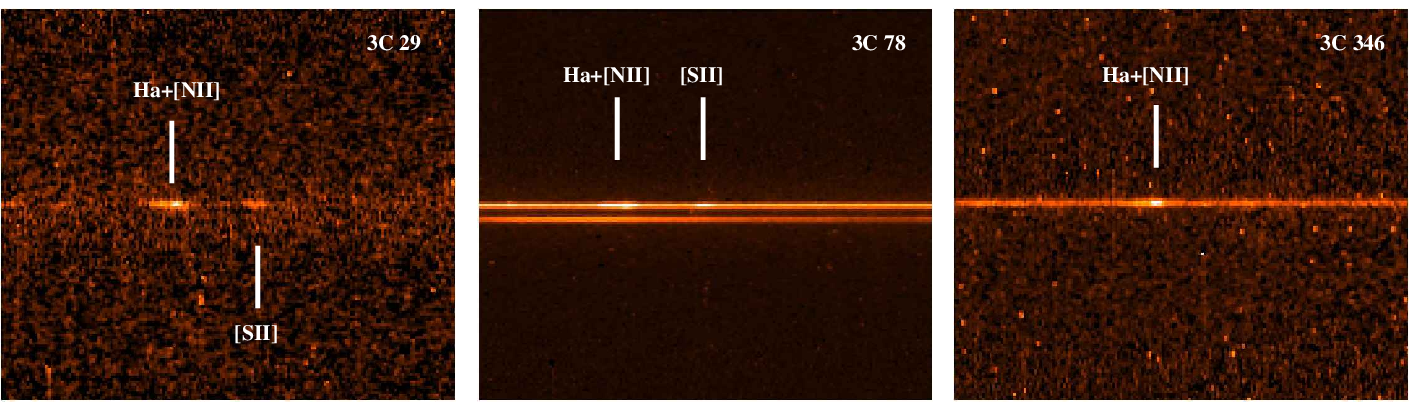}}
\caption{Continuum subtracted narrow-band images (upper six panels,
North points up) 
or 2D long slit spectra (lower three panels) for nine 3C
sources not in common with the VK02 sample. In all cases the images 
are dominated by the presence of a compact emission line region. 
The three spectra show that line emission is spatially unresolved.
for 3C~83.1 we present the narrow band image, without continuum
subtraction, see text.} 
\label{celrpics}
\end{figure*}

The narrow-band,  continuum subtracted\footnote{For 3C~83.1  the broad
band  image is affected by  the  diffraction spikes  of a  bright
nearby star. Therefore, we cannot determine the continuum subtraction
outside the nucleus.}  images (or
the  2-D  spectra   when  no  images  are  available)   are  shown  in
Fig. \ref{celrpics}  revealing the presence of a  clearly defined CELR
in all 9 objects.

We performed the measurements of  the CELR fluxes following a strategy
as similar as  possible to the one used for the  UGC sources.  For the
measurements from narrow-band imaging  we extracted the \Ha+[NII] flux
of  the line  component  within  an  aperture  of  radius
0\farcs1;  from  the  spectra,  we extracted  one-dimensional  nuclear
spectra by summing over three pixel rows, corresponding to $0\farcs3$,
centered on  the peak of the  emission through the  0\farcs2 slit.  We
then measured  the flux of  the H$\alpha$+[NII] emission  line complex
using the IRAF  task {\it splot}. For the 3  objects with both spectra
and images  the resulting H$\alpha$+[NII] agrees within  better than a
factor of 2. Spectral measurements are generally more accurate as they
have uncertainties ranging from 30  \% in the case  of 3C~28 to 
about 15  \% for NGC~6251
while those derived from the  images have uncertainties between 25 and
50 \%, overall similar to those of the VK02 sample.
 
\begin{table*}
\caption{Datasets used for this work}
\label{tabobs}
\centering
\begin{tabular}{l c c c c c}
\hline\hline
Name & Program ID & Data type & Instrument  & Filter/Grating & Exp. Time \\
\hline
3C~029    & GO8700 & spectrum  &  STIS-CCD  &  G750L  &  360s  \\
3C~78     & GO8700 & spectrum  &  STIS-CCD  &  G750L  &  720s \\
3C~83.1   & GO5957 & image     &  WFPC2     &  F673N  &  600s \\
3C~277.3  & GO8700 & spectrum  &  STIS-CCD  &  G750L  &  360s \\
          & GO5957 & image     &  WFPC2     &  FR680N  &  600s \\
3C~338    & GO8700 & spectrum  &  STIS-CCD  &  G750L  &  360s \\
          & GO5957 & image     &  WFPC2     &  F673N  &  600s \\
3C~346    & GO8700 & spectrum  &  STIS-CCD  &  G750L  &  360s \\
3C~442    & GO5927 & image     &  WFPC2     &  F673N  & 2500s \\
3C~465    & GO8700 & spectrum  &  STIS-CCD  &  G750L  &  360s \\
          & GO5957 & image     &  WFPC2     &  F673N  &  600s \\
NGC~6251  & GO6653 & image     &  WFPC2     &  F673N  & 5700s \\
\hline
\end{tabular}
\end{table*}

Before  we proceed,  there are  a  few points  worth discussing.   The
selection for  the observations  of 16 3CR  sources for which  we have
line data  are biased toward  a significantly lower redshift  than the
complete FR~I/3CR sample, being  limited almost exclusively to sources
with $z < 0.1$, with only 1  exception.  It must be noted that none of
the (few) 3C objects without  optical core detection has been observed
in  line  emission,  so  we  cannot  quantitatively  explore  possible
differences from nucleated  galaxies.  Conversely, several UGC sources
show a CELR where no optical  core is seen: most likely, isolating the
small spectral range of the  \Ha+[N II] complex enables us to increase
the  contrast against  the  host  galaxy with  respect  to broad  band
imaging and  to isolate the nuclear  emission at even  lower levels of
activity.   As no  optical,  radio or  emission-line  core fluxes  are
detected in UGC 6635, no optical  core was detected in UGC 8419 and in
UGC  11718, and  no  measurement for  the  radio core  of  UGC 7115  is
available, the number of representative points for the UGC sources can
vary in the different diagrams.

Results   from    broad   band   HST   imaging    are   presented   by
\citetalias{chiaberge:ccc} and \citetalias{gijs02}  for the 3C and UGC
samples    respectively.    Note   that    for    some   3C    sources
\citetalias{chiaberge:ccc}  the continuum broad  band fluxes  were not
corrected  for the  line  contribution as  this  information was  not
available.  The  measurements of  the  CELR  fluxes (corresponding  to
equivalent  widths  between  200  and  600 \AA)  confirm  the  typical
fractional contribution between 10 and  30\% of the broad band fluxes,
a contamination negligible for our purposes.

From the above papers we take the values of the optical nuclear fluxes
as   well  as  the   radio  measurements   they  collected   from  the
literature.  Total radio  luminosity are  taken at  the  lowest energy
available (between  178 and 1400 MHz)  while radio cores  come in most
cases from 5 GHz VLBI data.  For consistency we converted optical core
luminosities  to  7020  \AA\   (adopting  when  necessary  a  spectral
index\footnote{We define the spectral index $\alpha$ with the spectrum
in the  form F$_\nu  \propto \nu^{-\alpha}$} of  1), and  the extended
radio luminosities to 178 MHz (adopting a spectral index of 0.7).

\section{The relationships between line emission and the radio/optical
properties}  
\label{results}  

In  this  Section  we  explore  the relationships  between  the  
line luminosities  with 
the optical  nuclear luminosity as  well as  both the
core  and  extended  radio  emission.  

\begin{figure} \resizebox{\hsize}{!}
{\includegraphics[width=\textwidth]{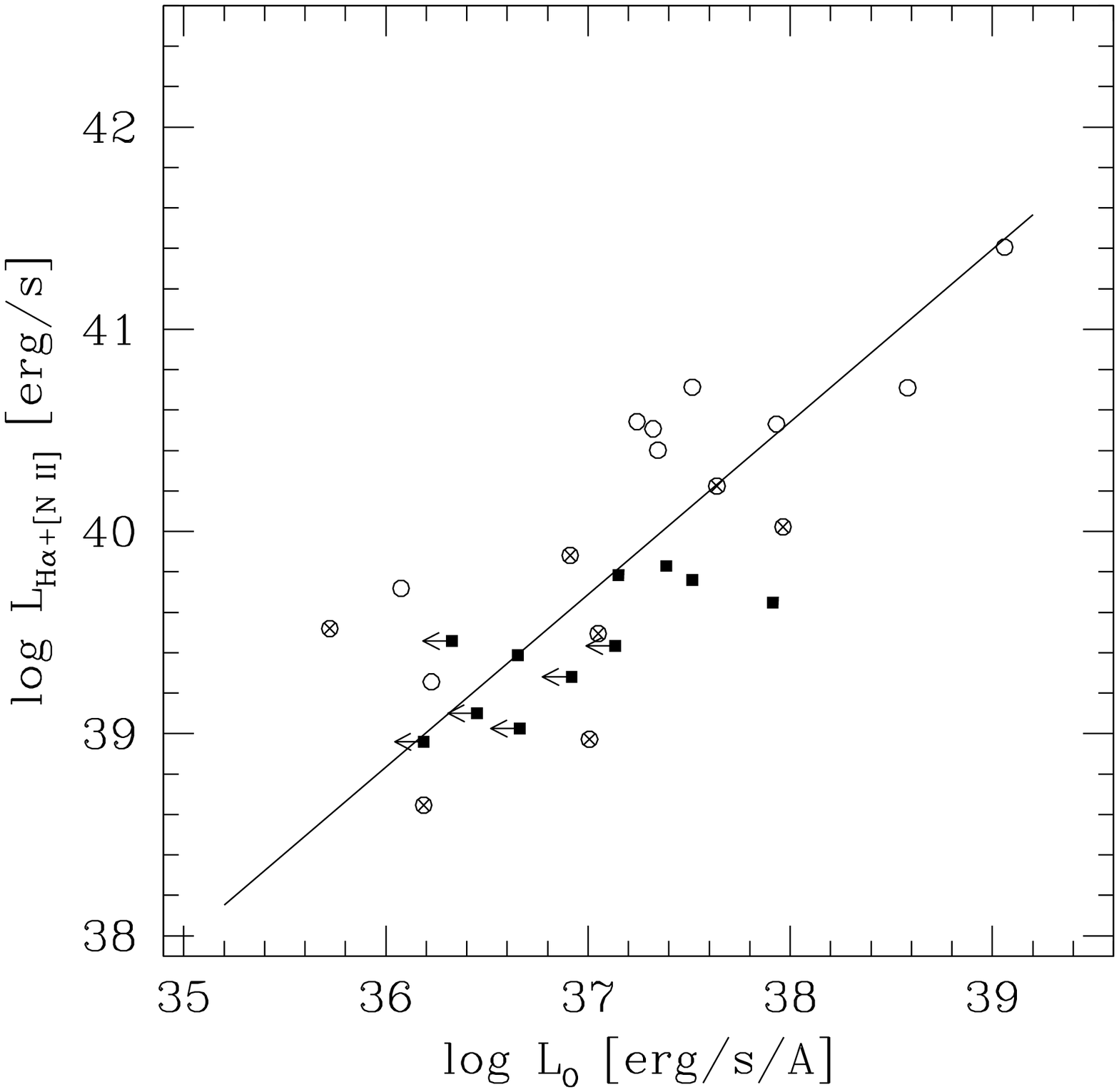}}
\resizebox{\hsize}{!}
{\includegraphics[width=\textwidth]{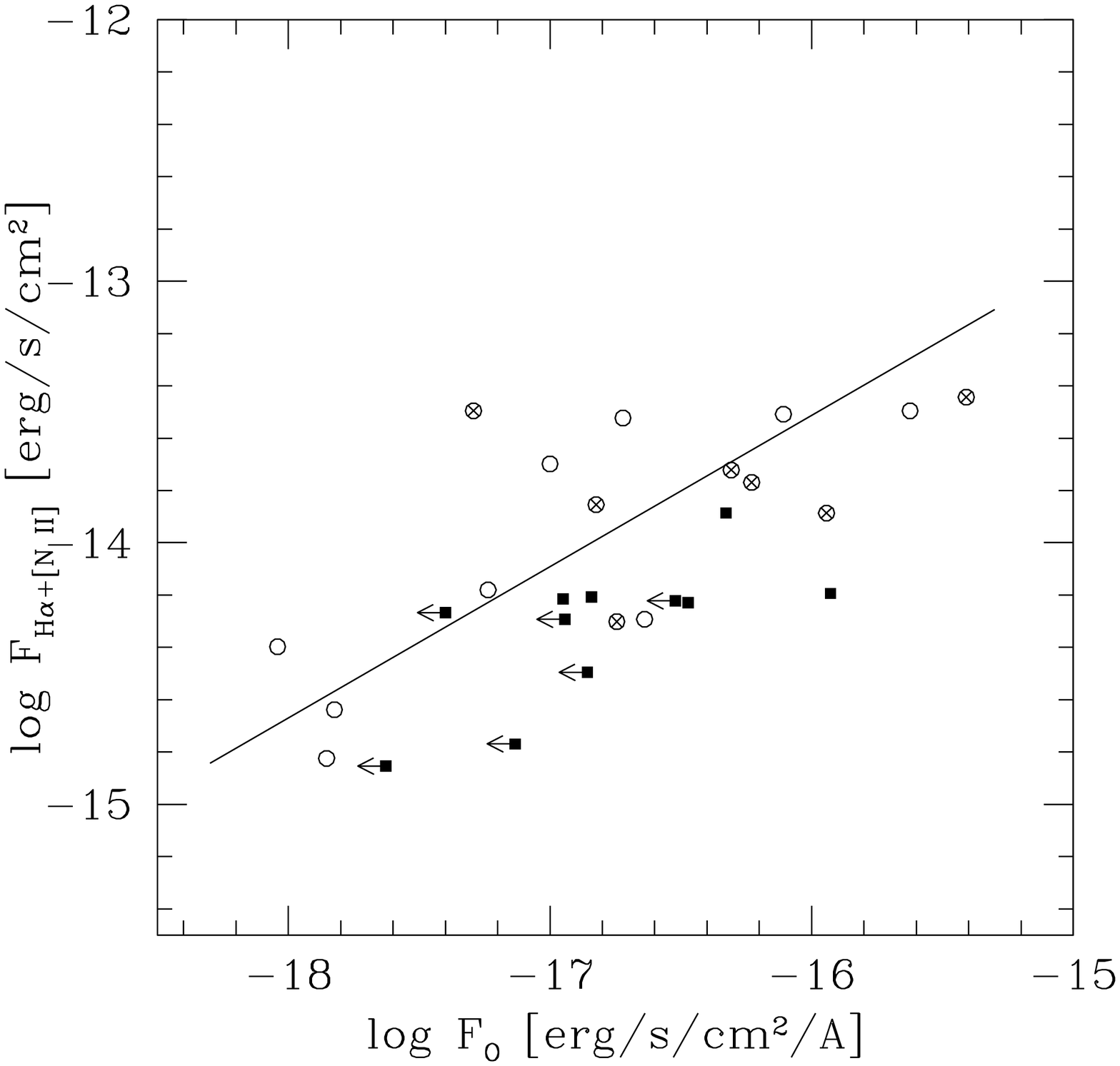}}
\caption{Relationship between emission line and optical continuum
luminosity and flux. Open circles are 3C sources while filled squares are UGC
galaxies. Objects common to the 3C and UGC samples are marked with
a crossed circle. The coefficients of the correlations 
have been estimated from the
3C sample alone, but UGC sample clearly follows the same relationship.}
\label{lohaall}
\end{figure}

In Fig. \ref{lohaall} we compare the \Ha+[N II]\ and nuclear optical
luminosities. A clear correlation is found 
with a correlation coefficient of 0.77. The slope of the correlation
(derived as the bisectrix of the linear fits using the
two quantities as independent variables) is m = 0.85$\pm0.13$ 
and the rms of the residuals is 0.45 dex.
Slope and intercept are derived from the 3C sources alone 
(as the bisectrix method cannot be used in presence of upper limits) but
UGC sources are clearly consistent with the same correlation.
A similar correlation (r = 0.71, m = 0.56 $\pm$ 0.18) 
is found also between line and continuum nuclear fluxes, suggesting
that the correlation found considering luminosities is not driven 
just by the common dependence on distance. To set this point on a more
quantitative ground we estimated the partial correlation coefficient between
line and optical nuclear emission excluding the dependence on
redshift\footnote{Given three variables,
if the correlation coefficient between $x_i$ and $x_j$
is $r_{ij}$, the partial correlation coefficient between two of them 
excluding the effect of the third variable is given by (see e.g.
Kendall \& Stuart 1979) 
$$
r_{12,3} = \frac{r_{12}-r_{13}r_{23}}{ \sqrt{1-r_{13}^2} \sqrt{1-r_{23}^2}} 
$$ }: $r$ is only marginally reduced to 0.75 which gives
a probability P = 8 $\times$ 10$^{-4}$ 
that these 16 points are taken from a random distribution.

The large range of redshift spanned by our objects (from 0.0037 to
0.162) raises the issue 
that the physical 
size of the extraction region used for the measurements of the line
emission varies substantially from object to object. 
To test the relevance of
this problem we repeated the analysis restricting to a sub-sample 
of 23 objects with $0.01 < z < 0.03$; although this reduces the range 
of line luminosities from 3.5 to 2.5 dex, 
the correlation between  L$_{H\alpha}+[N II]$  and $L_o$ is still
present with no significant changes of the fit parameters. 

Using ground based measurements collected by \citet{zirbel95} of line
intensity the results are also essentially unchanged. 
Restricting to the 9 sources common to the \citeauthor{zirbel95} and HST
samples the correlation has a slope of 0.82 $\pm$ 0.15
with the normalization a factor of $\sim$ 2.3 larger. 

\begin{figure} \resizebox{\hsize}{!}
{\includegraphics[width=\textwidth]{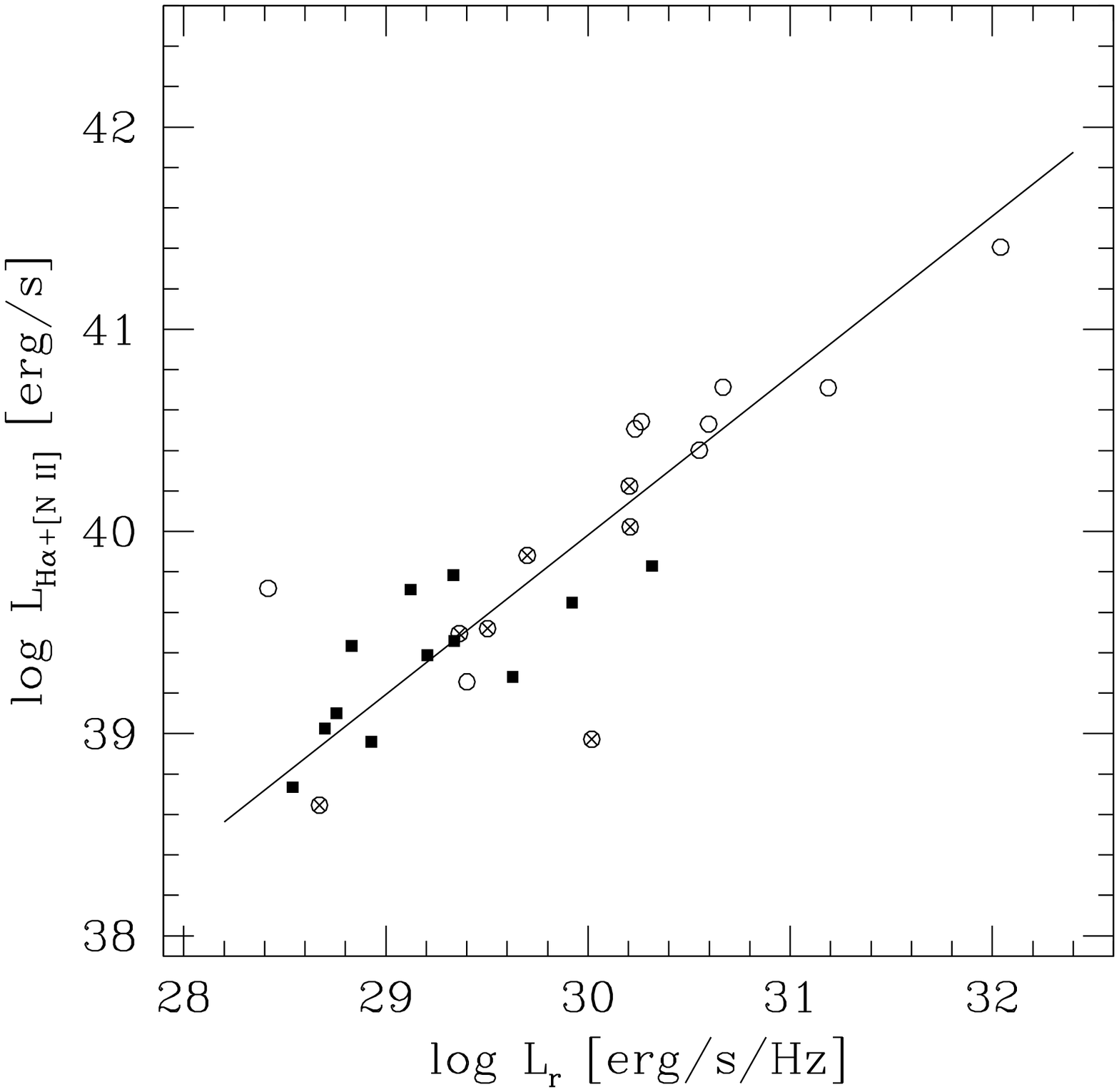}}           
\resizebox{\hsize}{!}
{\includegraphics[width=\textwidth]{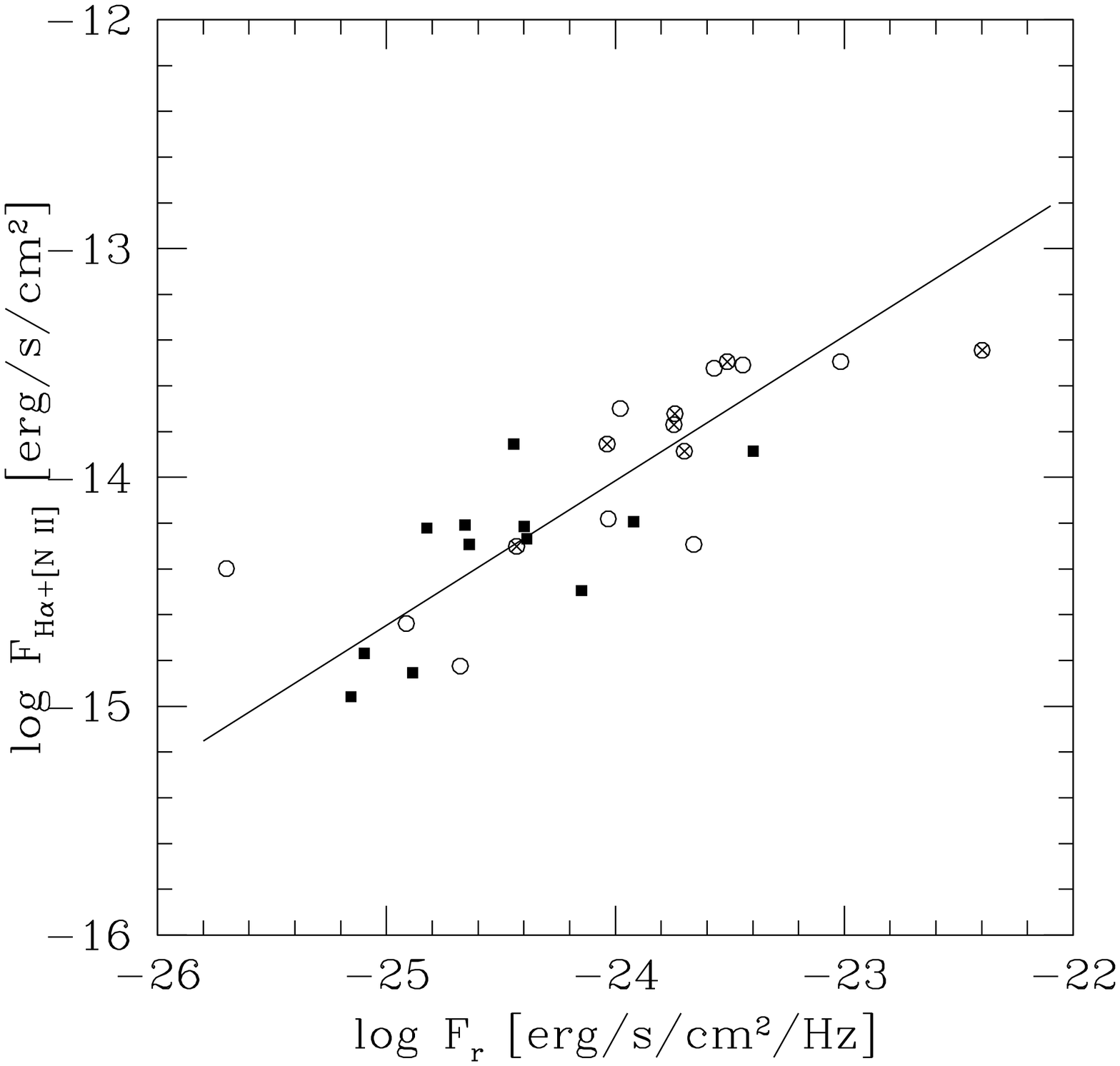}}
\caption{Relationship between emission line and radio core luminosity
and flux. Symbols as in Fig. \ref{lohaall}}
\label{lcorehaall}
\end{figure}

A similarly tight correlation is found comparing the radio core and
emission line luminosity, as already found by \citetalias{gijs02} for
the UGC sample, with a
correlation coefficient of 0.86 and a slope of 0.79$\pm0.10$ (see
Fig. \ref{lcorehaall}).
This result is somewhat expected on the base of the strong correlation between
optical and radio emission in the cores of FR~I radio-galaxies. 
This comparison is nonetheless interesting mainly due to the higher
completeness of detection of radio cores with respect to the
optical ones particularly for the UGC sample. This enables to see that 
the correlation holds also at the lowest levels of nuclear luminosity, 
smoothly extending
from the 3C to the UGC sample. The scatter of the correlation between 
core and line luminosity is slightly smaller
(0.39 vs 0.45) than for the optical core.
Again, the correlation between radio core and \Ha+[N II] is present
also considering fluxes instead of luminosities; the correlation
coefficient removing the effects of redshift is 0.79 
(P = 4 $\times$ 10$^{-7}$ for 29 points).

\begin{figure} \resizebox{\hsize}{!}
{\includegraphics[width=\textwidth]{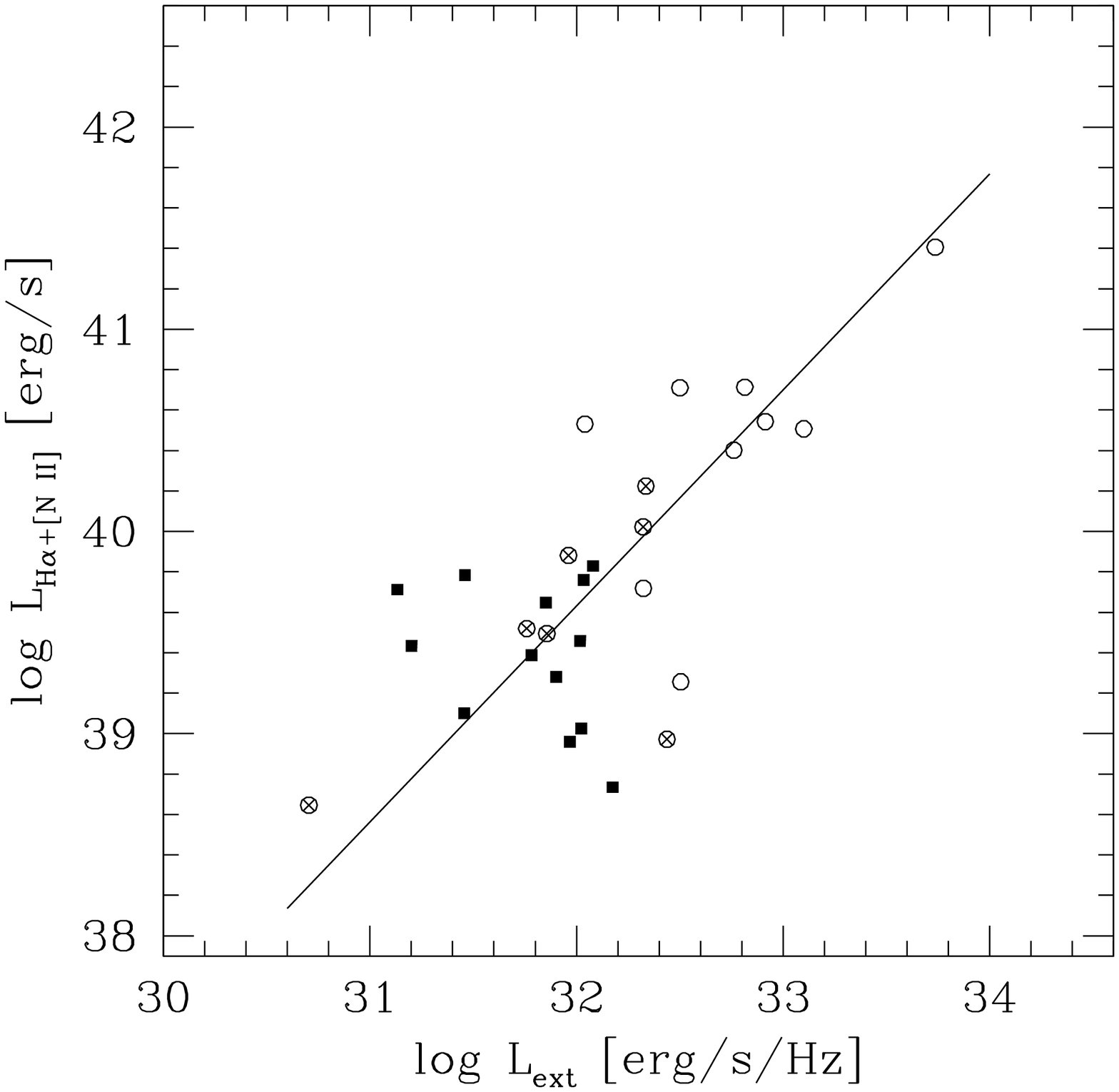}}
\resizebox{\hsize}{!}
{\includegraphics[width=\textwidth]{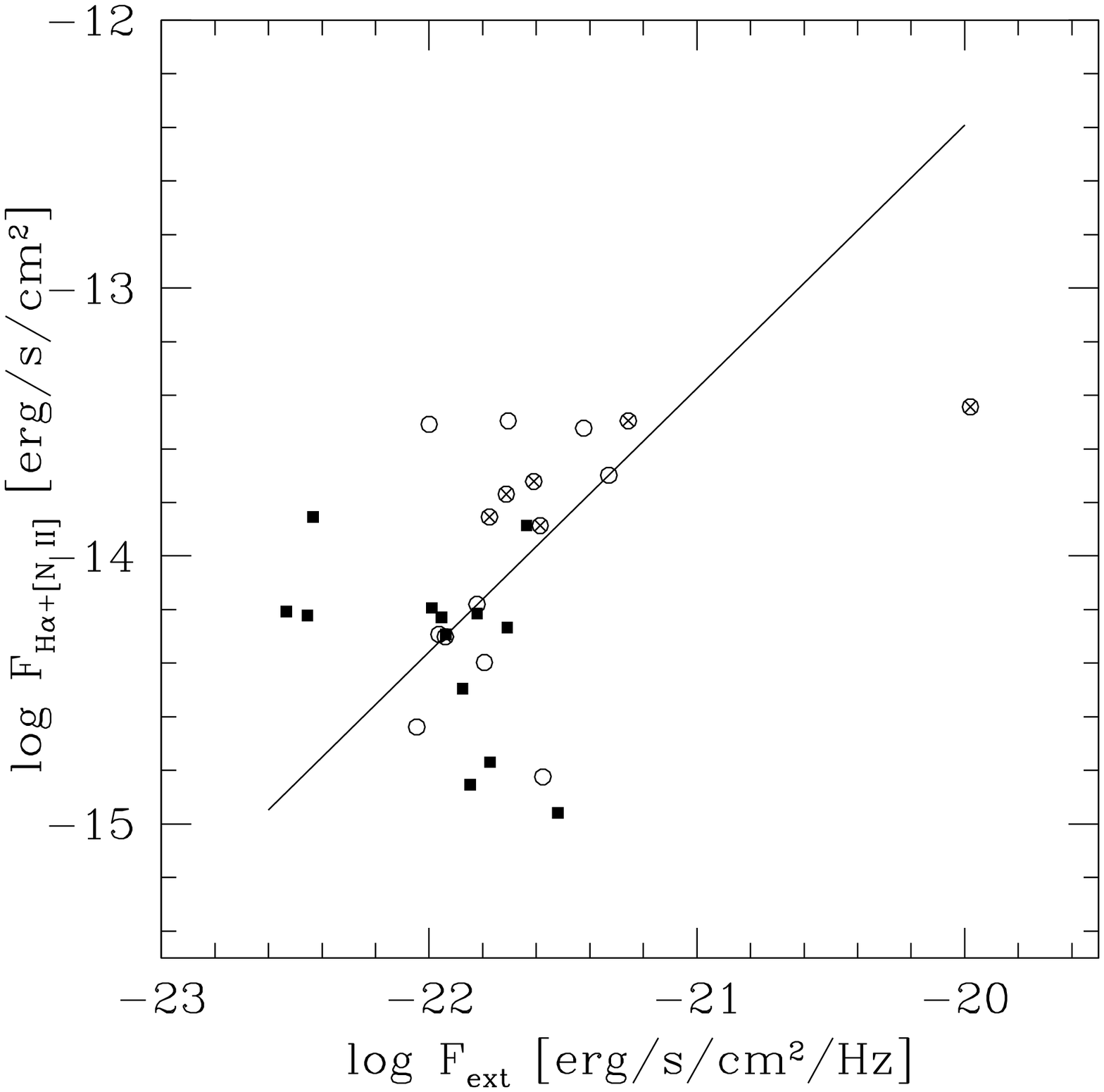}}
\caption{Relationship between emission line and extended radio
  luminosity
and flux. Symbols as in Fig. \ref{lohaall} }
\label{lexthaall}
\end{figure}

A correlation between  \Ha+[N II] and  extended  radio luminosity is also
present (see Fig. \ref{lexthaall}), but   
in this case the partial correlation coefficient is substantially
smaller, 0.29, indicating that it is mostly driven by their common
dependence on redshift.

\begin{table}
\caption{Correlations summary}
\begin{tabular}{l l c c c c } \hline
Var. A    & Var. B &  r$_{AB}$ & r$_{AB,z}$ & Slope & rms     \\
\hline								          
L$_{H\alpha}+[N II]$ & L$_{opt}$  & 0.77 & 0.75 & 0.85$\pm$0.13  & 0.45  \\
F$_{H\alpha}+[N II]$ & F$_{opt}$  & 0.71 &  --  & 0.56$\pm$0.18  & 0.41  \\
L$_{H\alpha}+[N II]$ & L$_{core}$ & 0.86 & 0.79 & 0.79$\pm$0.10  & 0.39  \\
F$_{H\alpha}+[N II]$ & F$_{core}$ & 0.81 &  --  & 0.63$\pm$0.13  & 0.36  \\
L$_{H\alpha}+[N II]$ & L$_{ext}$  & 0.69 & 0.29 & 1.07$\pm$0.14  & 0.46  \\
F$_{H\alpha}+[N II]$ & F$_{ext}$  & 0.40 &  --  & 0.98$\pm$0.18  & 0.41  \\
\hline
\end{tabular}
\label{tab0}

\end{table}

\section{The ionization mechanism of the CELR}
\label{mechanism}  

The correlation  between continuum and line luminosities  found in our
sample of  low luminosity radio-galaxies is reminiscent  of the well
known strong
connection observed in other classes of AGNs \citep[e.g.][]{adams75,netzer92} 
and similarly
suggests a  direct link between  the optical nuclear sources  with the
source of ionizing photons responsible  for the ionization of the CELR.
This idea must be tested quantitatively in order to establish 
if the higher energy photons (UV
and soft X-ray) associated to the optical sources provide a sufficient
photons flux to ionize the CELR. In the case of LLRG this 
approach requires particular care due to the paucity of 
information on their nuclear spectra and the possible effects of anisotropy.

{\subsection{The ionizing photons budget for the CELR}
For case B recombination, the flux of
Balmer line photons, $Q_{H\beta}$, is related to the flux of incident ionizing
photons, $Q_{ion}$,  
as $Q_{H\beta} = 0.12  Q_{ion}$\cf\ where \cf\ is the gas covering
factor \citep{osterbrock89}. 
Before applying the photons counting argument, we must
convert the ${H\beta}$ fluxes to our \Ha+[N II] measurements.
We scale the \Hb\ to  the \Ha\ luminosity  adopting the standard
3.1 ratio \citep[e.g.][]{gaskell84}. The typical
value of 4 for the  [N II]$\lambda\lambda$6584,6548/\Ha\  ratio measured
for  the UGC  sample \citep{noelstorr03}  provides a further 
factor of  5 correction. In the remaining of this
Section we use \Ha\ to refer to the \Ha+[N II] complex for sake of simplicity. 

We  start  with a  simple  approximation  for  the nuclear  properties
assuming i) a  power law extrapolation of the  optical core luminosity
with  a constant spectral  index $\alpha$,  i.e. $L_{\nu}  = C  (\nu /
\nu_{opt})^{-\alpha}$ and  ii) isotropy of the  nuclear emission.  The
effects of these  assumptions are discussed in detail  below.  In this
case, the  flux of ionizing  photons is obtained integrating  over the
nuclear  spectrum above  $\nu_0$, the  frequency corresponding  to the
ionization energy of hydrogen, i.e.

$$Q_{ion}= \int_{\nu_0}^\infty {{L_{\nu}} \over {h \nu}} d \nu = 
{C \over {\alpha h}} \left({{\nu_0} \over {\nu_{\rm
      opt}}}\right)^{-\alpha}
~~~~~~~~~~~~~~\rm{[photons ~~s^{-1}]}$$ 

The   spectral  indices  measured   by  \citet{donato04}   for  FR~I
radio-galaxies  from  the  optical  to  the X-ray  are  in  the  range
$\alpha_{\rm{OX}}\sim 0.8-1.2$  with an median value of  1, the latter
value  we adopted  for  the  integration of  the  nuclear spectrum  to
estimate the ionizing photons  flux. The corresponding model estimates
of  \Ha\ luminosity,  $L_{H\alpha  {\rm mod}}$,  are  compared to  the
observed  ones in  Fig.  \ref{model}.  For most  objects  the covering
factors of  the CELR (\cf), i.e.  the ratio between  the observed line
luminosity and the one predicted  by photon counting, are in the range
\cf $\sim$ 0.1 - 1 with a median value of 0.28.

Overall,  they compare  favorably with  the covering  factors  for the
Broad  Line Region (BLR)  estimated by,  e.g., \citet{kinney85}  for a
sample of nearby QSO as they  found a range of 0.04-0.47 with a median
of  0.17. For  the  Narrow Line  Region  (NLR) the  situation is  more
complex: as the luminosity of the narrow lines in both Seyfert and QSO
is at least  10 times fainter than the  corresponding broad components
\citep{boroson92,osterbrock82},  a  smaller  covering  factor,  a  few
10$^{-2}$,  might be expected.  However, this  might be  a substantial
underestimate  due  to   dust  embedded  in  the  NLR   (as  shown  by
\citealt{netzer93}).  Nonetheless, the relatively large values of \cf,
although acceptable  from a geometrical  point of view and  subject to
the  relatively  the large  uncertainties  typical  of this  approach,
warrant a more critical assessment of this result.

\begin{figure} \resizebox{\hsize}{!}
{\includegraphics[width=\textwidth]{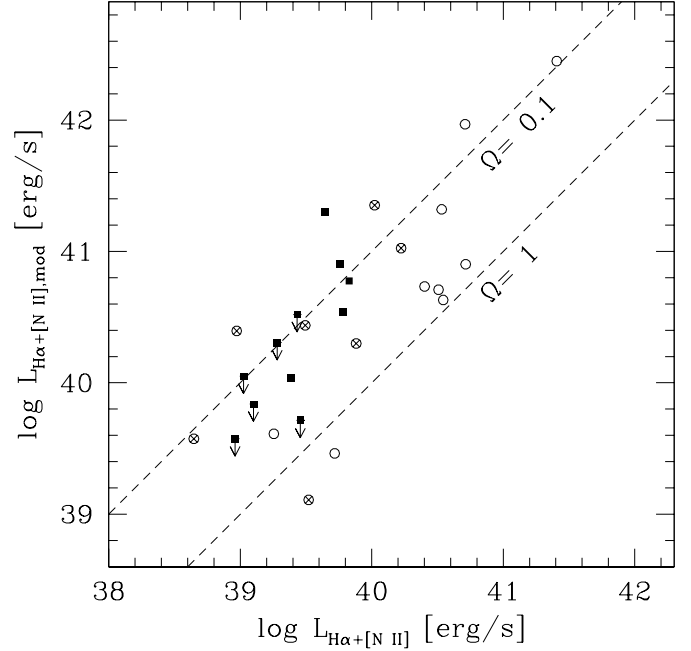}}           
\caption{\Ha\ luminosity estimated from the photons counting argument
assuming \cf =1, compared to the observed luminosity. The ratio between these
quantities provides the CELR covering factor, \cf. Lines of constant \cf\
are shown.}
\label{model}
\end{figure}

{\subsection{Photons budget and nuclear spectra}
We  then estimate how  the derived  line luminosity  $L_{H\alpha {\rm
mod}}$,  and  consequently   the  covering  factors,  change  adopting
different nuclear spectra. First of  all, still in the assumption of a
power law spectrum we note that the flux of ionizing photons scales 
with the spectra index as
$[\alpha  \, (\nu_0/\nu_{\rm  opt})^{\alpha}]^{-1}$. 
Thus, adopting  the extreme  observed value of  spectral index  of 0.8
(1.2), would reduce (increase) the covering factors by  a factor of 2.

To allow for a more realistic input spectrum we also used data for two
objects,  namely NGC  6251 and  Cen  A, for  which a  complete set  of
nuclear measurements exist over the full energy range, thus allowing a
detailed analysis of their  Spectral Energy Distribution (SED). SED of
these two objects have been recently studied by \citet{chiaberge:cena}
and  \citet{chiaberge:6251}  and shown  to  be  well  reproduced by  a
synchrotron self-Compton  model.  Here we  integrate the fit  to their
SED to  estimate the flux of  ionizing photons. The  advantage of this
method  is that  it allows a better estimate of the contribution  of  the non
thermal  nuclear emission  also  in spectral  regions unaccessible  to
observations, most notably  the far UV radiation to  which most of the
ionizing  photons  are   associated.   The  effective  spectral  index
(i.e. the  power law index  $\alpha_{\rm{OX}}$ that provides  the same
number of ionizing photons of the model  SED) is 1.6 for Cen A and 1.3
for NGC 6251 respectively.  These  values are slightly larger than the
measured $\alpha_{\rm{OX}}$  due to  the fact that  the SED in  the UV
region   is    even   steeper   that   predicted    by   the   average
$\alpha_{\rm{OX}}$.  
A  similar indication  for a  rather steep  spectral shape  in  the UV
regime comes from the  HST observations of radio-galaxies discussed by
\citet{chiaberge:uv}.  The  lowest value  of  spectral index  measured
between 2500  and 7000  \AA\ for FR~I  radio-galaxies is in  fact 1.2.

{\subsection{Photons budget and anisotropy}
Discussion of  the assumption of  isotropy of the nuclear  emission is
clearly another very important issue.  In fact, as the radio and (most
likely)  the   optical  emission  originates   from  the  base   of  a
relativistic jet, a dependence  on orientation of the nuclear emission
is  expected.   In  particular,  evidence for  a  substantial  Doppler
boosting  in   the  optical core   of  radio-galaxies  was   found  by
\citetalias{chiaberge:ccc}  comparing   the  optical  and   extended  radio
luminosity of  the 3C/FR~I: objects seen  closer to the  jet axis have
brighter nuclei for a given radio luminosity.  
Taking into account anisotropy is
then essential for a proper ionizing photon budget.

In this  context, as already  noted by \citetalias{gijs02}, 
the  relationship between
line  and  continuum  luminosity  contains  two  important  pieces  of
information on  this issue. Contrarily  to the optical  nuclear light,
line emission is not affected  by beaming and is most likely isotropic. 
An object  seen at
increasingly smaller angles from the  jet axis increases its nuclear
flux  at  constant  line  luminosity.  This  should  correspond  to  a
connection between  orientation and the  location of each  object with
respect to the $L_{\rm o}/L_{\rm H\alpha}$ correlation\footnote{
This trend can also be described as a decrease of the lines equivalent
width at decreasing angles between the jet axis and the line of sight.
The very low equivalent widths defining BL Lacs, 
the most closely aligned objects according to the FR~I/BL Lac unified model, 
would represent the extreme manifestation of this effect.}.

To test this idea, we used  as indicator of orientation the radio core
dominance,  i.e.  the  ratio  between radio  core  and extended  radio
luminosity $R = L_{\rm core}/L_{\rm  ext}$. This ratio has been widely
used in  the literature  to derive the  orientation of radio  loud AGN
despite the relatively  large scatter and the known  dependence of $R$
on radio  power.  Nonetheless,  Fig. \ref{orient7} shows  the expected
trend: objects with larger core  dominance value (seen closer to their
jet axis) show a larger  negative residuals from the $L_{\rm o}/L_{\rm
H\alpha}$ correlation,  i.e. they  have optical sources  brighter than
average at given emission line luminosity. 
More quantitatively, with a correlation coefficient of 0.49, 
the probability that the 20 detections 
are taken from a random distribution is P = 0.028.
As continuum and line fluxes are measured essentially at the
same wavelengths, this conclusion is independent of dust absorption.

This  result confirms,  first of  all, that  optical cores  are indeed
significantly  affected by  relativistic  beaming.  Furthermore,  this
trend shows that at least part of the scatter of the $L_{\rm o}/L_{\rm
H\alpha}$  correlation  is   due  to  orientation,  strengthening  the
connection between line and core luminosity.

\begin{figure} \resizebox{\hsize}{!}
{\includegraphics[width=\textwidth]{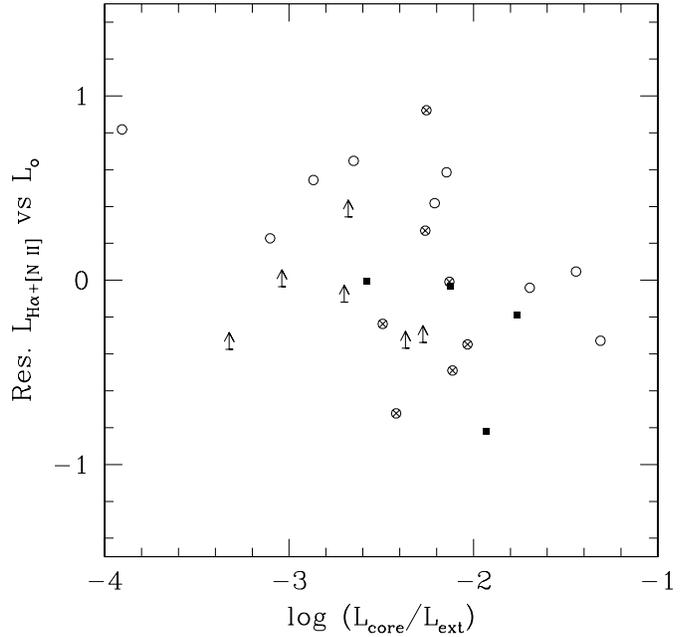}}           
\caption{Residuals of the L$_H{\alpha}$  vs L$_{\rm o}$ correlation
against the ratio between core and extended radio emission.}
\label{orient7}
\end{figure}

A  second important  result is that the  spread  of the  correlation
carries quantitative information  concerning the importance of Doppler
boosting.  Following  \citetalias{gijs02}, this  can  be  compared to  the
theoretical  expectations of  the dispersion  of a
randomly oriented  sample of  radio-galaxies. 
Here we  have sufficient
detection of optical cores to perform the analysis of the residuals in
this  band. Since Doppler amplification depends on the local
spectral index $\alpha$, optical data (for which $\alpha \sim 1$) 
set stronger limits on  the  effects of beaming than the radio
measurements (where $\alpha \sim 0$)\footnote{The  observed  flux scales  with
$\delta^{p+\alpha}$  (p=2 applies  to a  continuous jet  and p=3  to a
blobby  jet) where $\alpha$ is the local spectra index
and $\delta=[\Gamma(1-\beta ~cos\theta)]^{-1}$  is the
Doppler factor with $\theta$ the  angle formed by the  jet with the
line of sight.}. Comparing scatter and residuals of 
our $L_{\rm  o}/L_{\rm H\alpha}$  correlation with Fig. 6 by 
\citetalias{gijs02} yields a
jet bulk  Lorentz factor of $\Gamma \sim$ 2. 
Clearly, this value should be considered as an upper  limit 
since other sources of scatter in the correlation 
are very likely to be present. 

Thus, we conclude that 
relativistic  beaming affects the optical nuclei and 
it is then important for  a correct
photon  budget.  For  example the  objects with  the  largest covering
factors  can  now   be  understood  as  those  seen   at  the  largest
inclinations  with  the most  de-amplified  nuclei.  However,  beaming
effects within the small range  of $\Gamma$ allowed by the analysis do
not substantially  affect the photon budget for  complete samples.  In
fact, the  median value  of \cf\ corresponds  to an angle  between the
line of sight  and the jet axis of 60$^\circ$ (the  median value for a
randomly distributed population).  At this angle 
the Doppler factor is
constrained (for $\Gamma$ \ltsima 2) to  
$\delta =  [\Gamma(1-0.5  \beta)]^{-1} $\ltsima 0.88  corresponding to 
a (de)amplification due to Doppler beaming 
($\delta^{p+\alpha}$ with p=2 and $\alpha$=1)  
of \ltsima 0.69 (and to a decrease of \cf\ by the same factor).

\bigskip
Summarizing,  the  extrapolation of  the  spectra  of optical  nuclear
sources  to  higher  energies  corresponds  to a  sufficient  flux  of
ionizing  photons  to support  the  observed  line  emission, but  the
re-processing process  into line photons must be  quite efficient, with
relatively high CELR covering factors. Anisotropy of the optical nuclei
due to  Doppler beaming is present  and accounts for at  least part of
the already small scatter between continuum and line luminosities, but
the photon budget estimate is  rather robust against this effect.  The
most effective  possibility to substantially  reduce \cf\ relies  on a
flattening of the spectral slope  toward the UV. But this is not expected
from the measured optical/X-ray spectral index, it is not observed at
least down to  the near UV (2500  \AA) and it is not  predicted by the
spectral fitting of the two best studied objects.

We conclude that a large CELR covering factor is an unavoidable
condition for the viability of the scenario in which the source of 
photons ionizing the CELR is directly associated to the optical
nuclear sources. 

\section{A physical picture of the compact ELR in radio-galaxies}  
\label{physics}  

HST  images and  spectra of  the  CELR of  LLRG also  provide us  with
important  information concerning  their physical  structure.   In the
following  we focus  on 3C~274  (M~87),  the closer  and best  studied
object of the sample.

First  of all,  the size  of the  compact emitting  region in  M~87 is
constrained by the HST images to be smaller than $R_{CELR} < 0\farcs05
\sim  3$  pc  as  this  is unresolved  at  0\farcs1  resolution.   The
electrons density in its central emission line region has been derived
by \citet{macchetto:m87}  from HST/FOS spectra measuring  the ratio of
the  sulfur  lines  [S  II]$\lambda6731$/[S  II]$\lambda6717$.  Moving
toward the nucleus this ratio  decreases reaching $0.6 \pm 0.1$ at the
central aperture, still  far from the ratio corresponding  to the high
density limit  (0.42). This  is indicative of  an increase  of density
from a few hundreds to about $\sim 4300$ cm$^{-3}$ (with a factor of 2
accuracy) at the nucleus.

As discussed in  \citet{osterbrock89}, the mass of ionized  gas can be
estimated from  the observed Balmer  line luminosity\footnote{The mass
and the filling factor of ionized gas are given by
$$M={{m_p+0.1m_{He}}\over{n_e\alpha_{H\beta}^{eff}h\nu_{H\beta}}}
L_{H\beta} ~;~~~ \epsilon={M \over {(m_p+0.1m_{He}) V}}$$ 
where $V$ is the volume of the emitting region.
We adopted a temperature of 10$^4$ K, i.e. a recombination
coefficient  $\alpha_{H\beta}^{eff}=3.03~10^{-14}$  cm$^3$ s$^{-1}$.}.
For L$_{H\alpha+[N II]} = 2~  10^{39}$ erg/s and the same scaling used
above  to convert  to L$_{H\beta}$  we obtain  a mass  of $M  =  120 ~
(10^4/n_e) M_{\sun}$.  Combining these  measurements with the limit to
the CELR volume, we derive the filling factor for  the ELR gas of
$\epsilon \sim 0.03$ for
the above measured  density. This value is actually  a lower limit (as
the measured ELR radius is an  upper limit) and it also indicates that
the size of the CELR  cannot be substantially smaller than $\sim$1 pc,
as the filling factor would exceed unity.

Limits on the CELR's size for the other more distant sources are
less stringent, but nonetheless they are constrained to be
smaller than $\sim 20$ pc for $\sim$ 80\% of the sample.  
Scaling the mass found for M~87 to the range of LLRG line 
luminosity yields a typical CELR mass of 10 - 10$^3 M\sun$. 

Another important information concerns the dynamics of the nuclear
gas. \citet{macchetto:m87} showed that the gas rotation curve 
can be accurately reproduced as Keplerian motions of gas arranged
in a thin disk around the central super-massive black hole.
This result is common to several other objects of our
3C sample, e.g. 3C~272.1 \citep{bower:m84}, NGC~6251
\citep{ferrarese:6251}, 3C~270 \citep{ferrarese:4261}, by
two UGC sources, UGC 11718 \citep[i.e. NGC~7052,][]{vandermarel:ngc7052} 
and UGC7455 \citep{gijs:4335} as well as in another
LLRG, Centaurus A \citep{marconi:cenabh}. 

Although clearly we do not have information on the dynamics
(or on the morphology) 
of the CELR, the Keplerian motion can be followed 
from $\sim$ 100 pc, down to the HST resolution limit, 
5 pc in the case of M~87. On the other hand,
as discussed above, the radius of its CELR cannot be substantially 
smaller than $\sim$1 pc. If the disk structure is preserved also
on this smaller scales,
the estimated covering factor of CELR requires a substantial
vertical thickness of the gaseous disk: for example, a value of \cf $\sim 0.3$
corresponds to an angular thickness of $\theta \sim 15^{\circ}$.
In this sense, the central emission line regions would probably be better
described as ionized tori.
 
This raises the issue of the vertical support for the CELR.
While thermal motions are negligible in this context, 
detailed model fitting of circumnuclear disks around black holes show that 
the measured velocity dispersions are generally in excess 
of what it is predicted from just spatial smearing (e.g. \citealt{gijs00},
\citealt{vandermarel:ngc7052}, \citealt{barth01}). 
This requires the
presence of an internal velocity component in addition to thermal
and rotational motions, possibly due to turbulence, that 
reaches in the innermost regions values of several hundreds
of km/s, of the same order of the rotation velocity at these radii.
While this line of argument does not answer the question of the
physical origin of the vertical support, it shows that this can be provided 
by an expected component of turbulent velocity. 

\section{Line emission vs radio-luminosity}
\label{line-vs-radio}  

Line  and radio luminosity  (both core and total radio power)
are known to correlate for the powerful FR~II radio-galaxies
\citep[e.g.][]{rawlings91,zirbel95,willott99}.

Results  presented in Sect.\ref{results} show a correlation
between L$_{core}$ and L$_{H\alpha}+[N II]$; this  
contrasts with previous ground-based determination of these relationships
by \citet{zirbel95} as they found 
a much flatter slope (0.30$\pm$0.12)
with a  larger dispersion, 0.79  dex. 
The most likely explanation for this discrepancy is the contamination by
the host galaxy of the  emission line luminosity, particularly for the
faintest objects. In fact, FR~I  appears to be smoothly connected with
non  active galaxies  as far  as their  line and  radio  luminosity is
concerned  and  they also show  a  correlation  
between NLR  and  host  galaxy's
luminosity.  Only in the brightest FR~I does the AGN component provide a
substantial contribution to the total line emission and induces the somewhat
loose trend they observed. By solely concentrating  on a well defined sample 
of relatively bright FR~I and
isolating the central emission  line component, a clearer view
of the connection between these quantities emerges.

The interpretation of the radio-core/optical correlation observed in our
sample of LLRG FR~I appears to be 
rather straightforward  in the scenario presented in the
previous sections. The line  luminosity correlates
with the optical  continuum as this is directly associated with the  source of
ionizing photons;  the optical continuum is most likely 
of synchrotron origin and
it is  produced by the same  mechanism as the radio  core.
  
\section{Discussion}  
\label{discussion} 

Broad-band HST images of LLRG revealed the presence of a correlation
between fluxes and luminosities of the nuclear optical sources 
with the corresponding radio cores, indicating a dominant non-thermal
(jet) origin for their optical emission. Narrow-band images
(and spectra) of LLRG show a strong correlation also between fluxes and
luminosities of the optical sources with those of their
compact emission line regions. A photon budget argument supports the idea
that these non-thermal nuclei provide a
sufficient flux of ionizing photons, i.e. that we are seeing a jet-ionized
CELR. 

The origin of the 
lack of prominent broad lines  in LLRG must also be revised in the
light of the present results.  In fact high density clouds in presence
of the ionizing  photon fields required to illuminate  the compact ELR
(regardless on their origin)  should correspond to a significant broad
lines flux. With a simple  scaling argument, adopting a BLR density of
$n_e  =  10^9 {\rm cm}^{-3}$,  the  BLR  luminosity  would exceed  the  CELR
luminosity     as      soon     as     $M_{BLR}/M_{CELR}     $\gtsima$
n_{e,CELR}/n_{e,BLR}$,  assuming  that  both structures  have  similar
covering factors.  This sets a  limit of $M_{BLR} < 10^{-2} M_{\sun}$,
several orders of magnitude below  the BLR mass estimated for luminous
AGN (see \citealt{baldwin03} for a detailed discussion on BLR masses).
We conclude that  in LLRG the dense clouds  commonly associated to the
BLR  are  virtually not  present.   A  possible  explanation for  this
finding has  been proposed  by \citet{laor03}: given  the relationship
between BLR radius, R$_{BLR}$, and nuclear luminosity, R$_{BLR}$ would
be smaller than  the minimum radius for clouds  survival against tidal
disruption  below  a  bolometric   luminosity  of  $L  \sim  10^{41.8}
(M_{BH}/10^8 M\sun)^2$, a condition met by our LLRG.

The low masses of both ELR  and BLR represent a further evidence for a
general paucity of  gas in the innermost regions of  LLRG, to be added
to the low  accretion rate (derived from the  low nuclear luminosity),
to the failure  of finding maser emission \citep{henkel98}  and to the
general lack of obscuring molecular tori.

The available information indicates that the CELR must have a
substantial thickness to provide the necessary covering factors.
If the disky structure of the emission line region observed at larger
radii is preserved down to the pc-scale, 
the CELR would be better described as ionized tori. 
This raises the question of how
these structures would compare  with the molecular tori of  other classes of
AGNs.   The   low  fraction   of  possibly  optically   obscured  LLRG
\citepalias[\ltsima 15\%,][]{chiaberge:ccc}  suggests that little dust
is   associated  to  the CELR.   More   quantitatively,  HST
ultraviolet observations \citep{chiaberge:uv}  provided evidence for a
relationship between UV absorption  and orientation as expected from a
flattened  absorber.  For  the sources  seen close  to  the equatorial
plane its  optical depth corresponds  to A$_V \sim 3$.   The arguments
presented in Sect.~\ref{physics} enable us to estimate also their {\sl
ionized} gas  column density  that can be  compared with  results from
X-ray observations.  The average density of ionized gas derived from a
spherical distribution is $<n_e> \sim 10^2$ cm$^{-3}$ corresponding to
a  column  density  of  $N_H  \sim  <n_e>  ~  r_{CELR}  \sim  10^{21}$
cm$^{-2}$.  This  represents an average value for  a randomly oriented
class  of objects,  but if the absorbing material is  partially
flattened, we should be seeing a fraction of LLRG along line of sights
not intercepting the  torus, while for the remaining  sources a larger
$N_H$ should be measured. This  scenario is indeed supported by recent
X-ray   observations   of   LLRG   providing  examples   of   moderate
photoelectric  absorption  \citep[e.g.][]{sambruna:3c218,sambruna:270}
with typical column density of a few $10^{22}$ cm$^{-2}$ contrasted by
the  majority of  objects  in which  no  absorption in  excess to  the
Galactic                component               is               found
\citep[e.g.][]{hardcastle:66b,hardcastle:31}.  In  this sense the most
significant difference between the putative  tori in LLRG and those of
Seyfert would be the  much lower  optical depth:  in fact  $\sim$ 75  \% of
Seyfert  have $N_H  > 10^{23}$  cm$^{-2}$ and  about half  are Compton
thick  with   $N_H  >  10^{24}$   cm$^{-2}$  \citep{risaliti99}.   The
resulting effects of the circumnuclear gas in LLRG  would be
radically different  from those of  the standard molecular  tori. Not only 
it allows a  direct view  of the  nucleus, but it
provides a preferred environment for the production of line emission.

It would be  also interesting at this point to  establish if a 
CELR is present  also in  other AGNs,  as this  would carry
important  information on  their  nuclear structure. 
Most likely it can be directly
viewed only  in type  I sources. Indeed,  the excess  of high-ionization
coronal  lines, such  as  [Fe  VII], in  Sy  1 with  respect  to Sy  2
\citep{murayama98} has been interpreted as an indication that they are
produced (for a large fraction of the total flux) at the inner wall of
the dusty torus; this is in close analogy with the situation envisaged
to explain  the properties  of LLRG, albeit  in a different  regime of
photons and gas density.

Despite  the internal  consistency of  the jet  ionized  CELR scenario
alternative interpretations, such  a standard disk ionization, clearly
can not  be excluded at  this stage.  However, in the latter case
the  strong correlation 
between the optical core, radio core and CELR luminosity,
would imply a close link
between  the  thermal radiation  from  the  disk  and the  non-thermal
radiation from the  jet. While a general link between these two
processes might be expected \citep[e.g.][]{falcke95},
such a close connection appears  contrived, but cannot be
discarded at face value. 

A further hint on the nature  of the ionization source in LLRG comes
from their optical spectra, which are invariably of the LINER type. It
has  been  suggested  that a  LINER  spectrum  can  arise from  a  low
ionization  parameter  
\citep[U$=Q_{ion}/(4 \pi r^2 n_e c) \sim    10^{-3}     -    10^{-4}$
e.g.][]{ferland83}.  However,  estimates for M~87 based  on the values
presented  in Sect.~\ref{physics}  sets it  at a  substantially higher
value as we find U $\sim$ 0.008 \cf$^{-1}$ for a CELR radius of 3 pc.  As an
alternative, \citet{nagao02} recently  showed that spectra harder than
those emitted by standard optically thick  accretion disks, produces
larger partially  ionized regions,  increasing the relative  weight of
low  ionization lines.  Clearly,  a detailed  spectroscopic study  and
modeling of CELRs is needed to investigate this issue.

\section{Summary and conclusions}  
\label{conclusions} 

With the  aim of  studying the origin  and properties of  the emission
line region of Low Luminosity Radio Galaxies, we collected HST spectra
and narrow  band images for  a total of  30 sources.  HST  images show
that a compact emission line region (CELR)  is present at
the center of all objects  studied, with just one exception.  Focusing
on this central compact  component, we minimize the contamination from
line  emission  not   directly  related  to  the  AGN,   that  can  be
significant in these low luminosity  objects. A clearer  view  of  the LLRG 
properties emerges by isolating the
central  emission  line  component.

The line  luminosities show  a strong  connection with
other properties  of these objects, most importantly 
it correlates over almost  four
orders of magnitude with the optical
nuclear  luminosity.  This trend is reminiscent of what is observed in
other classes of more powerful 
AGNs and similarly suggests a direct link between the
optical nuclear  sources and the  source of the photons  ionizing the
NLR.  The extrapolation  of the spectra of optical  nuclear sources to
higher energies, corresponds to a sufficient flux of  ionizing photons, but
requires relatively high covering factors with a median value of 0.28.
We tested both the effects of different choices for the nuclear
spectra as well as of anisotropy of the  optical 
nuclei. Beaming effects are present but 
within  the small range allowed (the  bulk jet Lorentz
factor is  constrained to be  $\Gamma$ \ltsima 2) do  not substantially
affect the photon budget.

HST  images and  spectra of  the CELR of  LLRG also  provide us  with
information  concerning  their physical  structure.   Using
observations of 3C~274  (M~87) 
we derive the following fiducial values: a limit to the NLR size
of $\sim  3$ pc,  a density  of $\sim 4000$  cm$^{-3}$, a  mass 10$^2$
M$\sun$ and a filling factor larger than $\epsilon \sim$ 0.03. Furthermore,
the  dynamics of  the  nuclear gas in LLRG is 
accurately  reproduced   as  Keplerian  motions   around  the  central
super-massive black hole  in many objects of our  sample down to a
scale of a few parsec. 
If the disky structure of the emission line region observed at larger
radii is preserved down to the pc-scale, the covering factor 
derived from the photon-budget can
be  accounted if the CELR has an angular
thickness   of at least $\theta  \sim   15^{\circ}$. 
In this sense the CELR would be better described as ionized tori. 
Indeed, detailed   modeling  of
circumnuclear  disks around  black holes often requires the  presence  of a
turbulent  velocity  component  of  the  same order  of  the  rotation
velocity, providing the disk's vertical support.

Our conclusions on the physical structure
of the CELR do not depend on the origin of the ionizing photons.
In particular we find that its mass is as small as 10 - 10$^3 M_{\sun}$
while any BLR mass must be smaller than $M_{BLR} <  10^{-2} M_{\sun}$,
i.e. the dense clouds commonly associated to the BLR are virtually not present.
When considered together with the low accretion rate and to the
general lack of obscuring molecular tori, these results 
represent a further evidence that the different behaviour of LLRG
from the overall AGN population can be ascribed to a general
paucity of gas in their innermost regions.

The internal  consistency of  the jet-ionized  CELR scenario
clearly cannot be used to rule out alternative interpretations,
although the  strong correlation of optical core vs line luminosity
requires a similarly strong connection between any other sources of
ionizing photons with the jet emission, which appears to be quite
contrived.

The CELR luminosity also shows a quasi-linear correlation
with the radio core, 
contrasting with a much flatter slope and larger dispersion of
previous determination of this relationship.
The most likely explanation for these discrepancies is the contamination of
the  emission line luminosity by a non-AGN extended component.
Concentrating  on a well defined sample 
of relatively bright FR~I and
isolating the central emission  line, a clearer view
of the connection between these quantities emerges.
The interpretation of the  radio-cre/line correlations observed in FR~I
is rather straightforward  in the proposed scenario of a jet
illuminated NLR, as being the result of the chain of correlations
linking line emission, optical cores and radio cores.

\begin{table*}
\caption{Properties of the 3C sample}
\begin{tabular}{l c c c c r r r r } \hline
         & UGC name & VK02 & NGC name &    z     &   F$_{r}$ &F$_{rext}$& F$_{opt}$   &F$_{H\alpha+[N II]}$    \\
\hline								          
3C~029   &   595 &     &      &  0.0448   &     93  &     15.1 &    5.8e-18  &   6.6e-15  \\
3C~031   &   689 & yes &  383 &  0.0169   &     92  &     16.8 &    1.5e-17  &   1.4e-14  \\
3C~066B  &  1841 & yes &      &  0.0215   &    182  &     24.6 &    4.9e-17  &   1.9e-14  \\
3C~078   &  2555 &     & 1218 &  0.0290   &    964  &     19.8 &    2.4e-16  &   3.2e-14  \\
3C~083.1 &  2651 &     & 1265 &  0.0251   &     21  &     26.6 &    1.4e-18  &   1.5e-15  \\
3C~264   &  6723 & yes & 3862 &  0.0206   &    200  &     26.0 &    1.1e-16  &   1.3e-14  \\
3C~270   &  7360 & yes & 4261 &  0.0074   &    308  &     55.5 &    5.1e-18  &   3.2e-14  \\
3C~272.1 &  7494 & yes & 4374 &  0.0037   &    180  &     19.4 &    5.9e-17  &   1.7e-14  \\
3C~274   &  7654 & yes & 4486 &  0.0037   &   4000  &   1050.0 &    3.9e-16  &   3.6e-14  \\
3C~277.3 &       &     &      &  0.0857   &     12  &      9.0 &    1.5e-18  &   2.3e-15  \\
3C~338   & 10409 &     & 6166 &  0.0303   &    105  &     46.9 &    1.0e-17  &   2.0e-14  \\
3C~346   &       &     &      &  0.1620   &    220  &     10.9 &    2.3e-17  &   5.1e-15  \\
3C~442   & 11958 &     &      &  0.0262   &      2  &     16.1 &    9.1e-19  &   4.0e-15  \\
3C~449   & 12064 & yes &      &  0.0181   &     37  &     11.5 &    1.8e-17  &   5.0e-15  \\
3C~465   & 12716 &     & 7720 &  0.0301   &    270  &     37.8 &    1.9e-17  &   3.0e-14  \\
NGC~6251 & 10501 &     & 6251 &  0.0240   &    360  &     10.0 &    7.8e-17  &   3.1e-14  \\
\hline
\end{tabular}
\label{tab1}

\medskip
F$_{r}$ radio core flux at 5 GHz in mJy, F$_{rext}$ total radio flux
at 178 MHz in Jy, F$_{opt}$ optical core flux at 7020\AA\ in erg/s/cm$^2$/\AA,
F$_{H\alpha+[N II]}$ flux of the compact emission line region in erg/s/cm$^2$. 
\end{table*}

\begin{table*}
\caption{Properties of the UGC sample}
\begin{tabular}{l c c r r r r } \hline
          & NGC name &    z     &   F$_{r}$ &F$_{rext}$& F$_{opt}$   &F$_{H\alpha+[N II]} $    \\
\hline								          
UGC~00408 &  193 & 0.0145  &   40 &  3.4  &1.1e-17   &   6.1e-15 \\
UGC~00597 &  315 & 0.0165  &  396 &  5.2  &4.7e-17   &   1.3e-14 \\ 
UGC~01004 &  541 & 0.0181  &    8 &  3.8  &$<$7.4e-18&   1.7e-15 \\
UGC~01413 &  741 & 0.0185  &   13 &  3.2  &$<$2.4e-18&   1.4e-15 \\
UGC~03695 & 2329 & 0.0191  &  117 &  2.3  &1.2e-16   &   6.4e-15 \\
UGC~05073 & 2892 & 0.0227  &   22 &  0.7  &1.4e-17   &   6.2e-15 \\
UGC~06635 & 3801 & 0.0109  &  --  &  3.4  &   --     &     --    \\
UGC~07115 &      & 0.0226  &  --  &  2.5  &3.4e-17   &   5.9e-15 \\
UGC~07455 & 4335 & 0.0154  &   15 &  7.9  &$<$3.0e-17&   6.0e-15 \\
UGC~08419 & 5127 & 0.0161  &    7 &  0.7  &   --     &   1.1e-15 \\
UGC~08433 & 5141 & 0.0177  &   71 &  3.0  &$<$1.4e-17&   3.2e-15 \\
UGC~09058 & 5490 & 0.0167  &   41 &  4.4  &$<$4.0e-18&   5.4e-15 \\
UGC~11718 & 7052 & 0.0139  &   36 &  0.8  &   --     &   1.4e-14 \\
UGC~12531 & 7626 & 0.0114  &   23 &  2.6  &$<$1.1e-17&   5.1e-15 \\
\hline
\end{tabular}
\label{tab2}

\medskip
F$_{r}$ radio core flux at 1.4 GHz in mJy, F$_{rext}$ total radio flux
at 408 MHz in Jy, F$_{opt}$ optical core flux at 8140\AA\ in erg/s/cm$^2$/\AA,
F$_{H\alpha+[N II]}$ flux of the compact emission line region in erg/s/cm$^2$. 
\end{table*}

\end{document}